\documentclass[a4paper,11pt]{article}
\usepackage{slashed}
\usepackage{pos}
\usepackage{amsmath}
\newcommand{\phibar}{\hskip1.0mm{\bar{\hskip-1.0mm\phi}}}
\newcommand{\phitilde}{\hskip1.0mm{\tilde{\hskip-1.0mm\phi}}}

\title{
\vspace*{-1.25cm}
\begin{minipage}{\textwidth}
\begin{flushright}
\texttt{\small
PoS(LATTICE2024)398  \\
DESY-25-021          \\
}
\end{flushright}
\end{minipage}\\[15pt]
\vspace*{+0.625cm}
        Absence of CP Violation in the Strong Interaction:\! Vacuum thwarts Axion}

\ShortTitle{Absence of Strong CP Violation}

\author{Gerrit Schierholz}

\affiliation{Deutsches Elektronen-Synchrotron DESY,\\
  Notkestra{\ss}e 85, 22607 Hamburg, Germany\\and\\II. Institut f\"ur Theoretische Physik, Universit\"at Hamburg,\\ Luruper Chaussee 149, 22761 Hamburg, Germany}

\emailAdd{gerrit.schierholz@desy.de}

\abstract{QCD admits a contribution to the action, the $\theta$ term, which potentially gives rise to nontrivial phases and violates CP. This is essentially a question of how the vacuum reacts to the $\theta$ term. In this talk I will address the problem using new developments on the lattice. The overall solution is contrasted with the axion `solution'.}

\FullConference{The 41st International Symposium on Lattice Field Theory (LATTICE2024)\\
 28 July - 3 August 2024\\
Liverpool, UK\\}

\begin{document}
\maketitle

\section{Introduction}

The strong CP problem is one of the biggest unsolved problems in the elementary particles field. Unlike most of the problems that demand new physics that goes beyond the Standard Model, the strong CP problem is primarily a problem with QCD itself. 

We consider Euclidean space-time. QCD allows for a CP-violating term, $S_\theta$, in the action,
\begin{equation}
  S = S_0 + S_\theta\,, \quad S_\theta = -\, i\, \theta\, Q \,
\end{equation}
called the $\theta$ term, where $\theta \in (-\pi,\pi]$ is the so-called vacuum angle and $Q$ is the topological charge. A finite value of $\theta$ is expected to result in an electric dipole moment of the neutron, which violates
CP and P. To date the most sensitive measurements of $d_n$ are compatible with zero. The current upper bound is $|d_n| < 1.8 \times 10^{-13}\, \textrm{e\, fm}$~\cite{Abel:2020pzs}. This is commonly said to mean that $\theta$ is anomalously small.
It goes on to say that if $\theta$ is taken to be zero, the theory has no enhanced symmetry, so it is natural to expect it to have a value of $O(1)$, which is the problem.

In this talk I will present a solution of the strong CP problem within QCD. The solution is rooted in the topological properties of the vacuum. This raises the question of how this fits in with the axion `solution'. The axion extension of the Standard Model~\cite{Peccei:1977hh}  not only eliminates any dependence on $\theta$, but is found to drastically change the long-distance properties of the theory.

\section{Preliminaries}

\begin{figure}[h!]
  \begin{center}
\includegraphics[width=5.5cm]{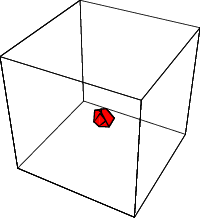} 
  \end{center}
  \vspace*{-0.25cm}
  \caption{The density of a typical zero mode in a given time slice on the $12^3\times 24$ lattice at lattice spacing $a=0.125 \, \textrm{fm}$. Shown is the isosurface $|u^\dagger u|=0.0005$ as given in~\cite{Koma:2005sw}. Results are from a cluster analysis~\cite{Ilgenfritz:2007xu}.}
  \label{fig1} 
\end{figure}

Lattice regularization in finite volume $V$ is assumed, although continuum notion is used occasionally. The topological charge is most conveniently expressed by
\begin{equation}
  Q = -\sum_{i} \int d^4x \; u^\dagger_i(x) \gamma_5 u_i(x)\,, 
  \label{top}
\end{equation}
making use of the Atiyah-Singer index theorem, where \{$u_i(x)$\} are the orthonormal zero-mode eigenvectors of the Dirac operator $\slashed{D}\, u_i(x) = 0$. On the lattice, the definition (\ref{top}) of the topological charge is realized by~\cite{Hasenfratz:1998ri} 
\begin{equation}
  Q = \frac{1}{2} \, \textrm{Tr}\,\gamma_5 D_N = -\, \frac{1}{2} \, \textrm{Tr}\,\gamma_5 (2-D_N) = - \sum_{\{\lambda=0\}} \, \left(u_\lambda, \gamma_5 u_\lambda\right) \,, 
  \label{topover}
\end{equation}
where $\{u_\lambda\}$ are the eigenvectors of the overlap Dirac operator, $D_N\, u_\lambda = 0$~\cite{Neuberger:1997fp}, with $n_+$ ($n_-$) eigenvectors having positive (negative) chirality, $\displaystyle \gamma_5 u_\lambda =\, {\raisebox{0.125cm}{+}\hspace*{-0.285cm}\raisebox{-0.125cm}{(-)}}\; u_\lambda$. Thus $Q = n_- - n_+  \in \mathbb{Z}$.
The definition (\ref{top}) of topological charge is preferred for several reasons. It is an integer from the start, and it does not need to be renormalized or stripped off its ultraviolet divergences, by means of the gradient flow for example. The zero mode eigenfunctions have been shown to be rather local~\cite{Koma:2005sw,Ilgenfritz:2006gc,Ilgenfritz:2007xu}. In Fig.~\ref{fig1} I show the density $u(x)^\dagger u(x)$ of a typical zero mode by its isosurface in a given time slice, taken from~\cite{Koma:2005sw}. It was already known that the eigenfunctions are strongly correlated with instantons of the appropriate charge~\cite{DeGrand:2000gq}. 

\begin{figure}[b!]
  \vspace*{-0.9cm}
  \begin{center}
    \includegraphics[width=10.5cm]{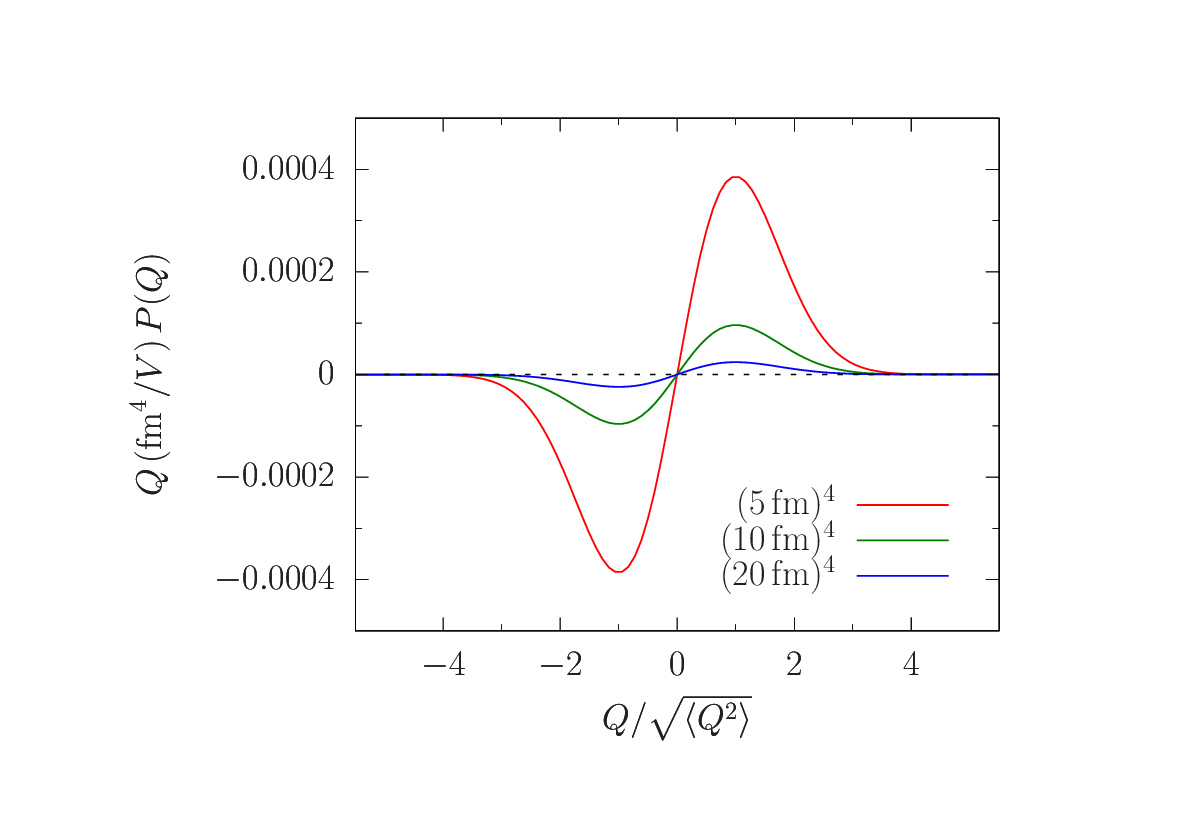}
  \end{center}
  \vspace*{-1.25cm}
  \caption{The fraction of zero modes in a subvolume of one fermi cubed as a function of $Q = \pm n$ for total volumes $V$ of $(5 \,\textrm{fm})^4$, $(10 \,\textrm{fm})^4$ and $(20 \,\textrm{fm})^4$. Scale setting is described in the text.}
  \label{fig2}
\end{figure}

A key result of the Dirac operator $\slashed{D}$ in two and four dimensions with generic (gauge) connections is that on a compact manifold with positive (negative) charge $Q$ is that it has exactly $n_-$ ($n_+$) zero modes, but never zero modes of both chiralities. In mathematical language $\textrm{dim ker}\, i\slashed{D} = |n_+ - n_-|$. This so-called `vanishing theorem' has been proven rigorously in~\cite{Ansourian:1977qe,Nielsen:1977aw,Maier}. On the lattice the absence of zero modes of different chiralities has been observed in several studies~\cite{Chiu:1998bh,Ilgenfritz:2007xu,Chiu:2011dz,DiGiacomo:2015eva,Chen:2022fid}. In the chiral effective theory~\cite{Leutwyler:1992yt} it arises naturally in the chiral limit. If we denote the total number of zero modes by $n$, then either $n = n_+$ or $n = n_-$. By all we know the topological susceptibility
\begin{equation}
  \chi_t = \frac{\langle Q^2 \rangle}{V} \equiv \frac{\langle n^2 \rangle}{V}
\end{equation}
is independent of the volume $V$. We may assume that the probability function $P(Q)$ for charge $Q$ is Gaussian,
\begin{equation}
  P(Q) \equiv \frac{Z_Q}{Z} =\frac{1}{\sqrt{2\pi \langle Q^2\rangle}} \;e^{-Q^2/2\langle Q^2\rangle}\,, \quad |Q|=n \,.
\end{equation}
Minor corrections should not matter. The density of zero modes is then given by
\begin{equation}
  \frac{\langle n \rangle}{V} = \sqrt{\dfrac{2}{\pi}} \, \dfrac{\sqrt{\langle Q^2\rangle}}{V} = \sqrt{\dfrac{2}{\pi}} \,\sqrt{\dfrac{\chi_t}{V}} \,,
  \label{density}
\end{equation}
which vanishes with the inverse square root of the volume. The fraction of zero modes to be expected in a subvolume of (say) one fermi cubed is given by $|Q|\, (\textrm{fm}^4/V) \, P(Q)$. The result is shown in Fig.~\ref{fig2} for $N_f=2+1$ quark flavors at the physical point, where $\chi_t=(79\,\textrm{MeV})^4$. Three volumes, $V = (5 \,\textrm{fm})^4$, $(10 \,\textrm{fm})^4$ and $(20 \,\textrm{fm})^4$ are considered. By increasing the lattice volume from $(5 \,\textrm{fm})^4$ to $(10 \,\textrm{fm})^4$ and $(20 \,\textrm{fm})^4$ the fraction of zero modes shrinks by a factor of $4$ and $16$, respectively. The convolution of two Gaussian distributions is again a Gaussian. Thus, the topological susceptibility on a sub-volume will be the same as on the total volume, $\chi_t = \langle Q^2\rangle/V = \langle Q^2\rangle_{\textrm{sub}}/V_{\textrm{sub}}$, where $\langle Q^2\rangle_{\textrm{sub}}$ can be computed, for example, by means of the `slab method' suggested in~\cite{Bietenholz:2015rsa}. 

\section{CP (non)violation: The electric dipole moment}

The key object of our studies is the partition function, which is the generating functional for correlation functions in the path integral formalism. The partition function is defined as
\begin{equation}
  Z(\theta) = \int\! \mathcal{D}[A_\mu]  \int [\cdots] \, \exp\{- i \theta Q - S_0 \}\,,
  \label{Z}
\end{equation}
where \raisebox{1.75pt}{$\,\int {\scriptstyle [\cdots]}$} stands for the integral over the fermion fields. The fermion fields are to be integrated out first, resulting in the fermion determinant, before the integrals over the gauge fields can actually be performed. In the continuum the gauge fields $A_\mu$ split into quantum mechanically disconnected sectors of topological charge. On the lattice this happens if the lattice spacing is sufficiently small, which we assume to be the case.
In the $\theta$ vacuum an $n$-point correlation function of operators $\mathcal{O}_1, \dots, \mathcal{O}_n$ then takes the form
\begin{equation}
  \langle \mathcal{O}_1\! \cdots \mathcal{O}_n \rangle_\theta = \langle e^{i\theta Q} \mathcal{O}_1\! \cdots \mathcal{O}_n \rangle = \frac{1}{Z(\theta)} \sum_Q \, \int_{Q}\! \mathcal{D}[A_\mu]\! \int [\cdots] \;\mathcal{O}_1\! \cdots \mathcal{O}_n \,\exp\{-i\theta Q - S_0\} \,.
  \label{op}
\end{equation}
The correlation function splits into two parts, a function that is odd in $\theta$ and one that is even. Only the odd part violates CP, while the even part is CP invariant. This means that even in the absence of CP violation the problem of naturalness mentioned at the start, suggesting that $\theta = O(1)$, continues to exist. 


The strong CP problem is largely a question of why the electric dipole moment of the neutron is so small. The dipole moment is given by
\begin{equation} 
  \vec{d}_n = \dfrac{\int d^3\vec{x} \: d^3\vec{y} \: e^{i \vec{p} \vec{x}}\, \langle N(\vec{x},x_0)\, \vec{y}\, J_0(\vec{y},y_0)\, \bar{N}(0) \rangle_\theta}{\int d^3x \: e^{i \vec{p} \vec{x}}\, \langle N(\vec{x},x_0) \, \bar{N}(0) \rangle_\theta}\,,
  \label{dipole1}
\end{equation}
where $N$ is the interpolating field of the neutron and $J_0$ is the time component of the electromagnetic current, with $x_0 \gg y_0 \gg 0$. The support of $\vec{y}$ is limited to the spatial size of the nucleon. Appropriate traces of the correlators are to be taken, which we have omitted. In practice, large ($\gg$) means about a fermi apart. Treating the $\theta$ term as a perturbation at first order, we obtain
\begin{equation} 
  \vec{d}_n = -\, i\,\theta \; \dfrac{\int d^3\vec{x} \: d^3\vec{y} \: e^{i \vec{p} \vec{x}}\, \langle \big(\sum_i \int d^4z\, u^\dagger_i(z)\gamma_5 u_i(z)\big)\,N(\vec{x},x_0)\, \vec{y}\, J_0(\vec{y},y_0)\, \bar{N}(0) \rangle}{\int d^3x \: e^{i \vec{p} \vec{x}}\, \langle N(\vec{x},x_0) \, \bar{N}(0) \rangle}\,,
  \label{dipole2}
\end{equation}
where the topological charge $Q$ in the numerator is given by the integral over the zero-mode eigenvectors (\ref{top}). It should be noted that the expression (\ref{dipole2}) is in full agreement with the schematic description based on the axial Ward identity given in~\cite{Guadagnoli:2002nm,Schierholz:2024var}, where the disconnected quark loop with $\gamma_5$ insertion is realized by the zero modes. 

The correlation function (\ref{dipole1}) extends over little more than the spatial size of the nucleon and spreads over Euclidean times of a few fermi only. A nonvanishing dipole moment arises from the interaction of the electromagnetic current with one of the zero modes (\ref{dipole2}). Given that the zero mode eigenfunctions extend over a fraction of a fermi only, the dipole moment (\ref{dipole2}) can be visualized by a fermion-line disconnected diagram of two rather local operators. The result will depend on the probability of finding a zero mode within the interacting range. If the zero modes are out of range, it does not matter whether they have positive or negative chirality, due to the absence of long-range interactions and the cluster decomposition property~\cite{Schierholz:2024var}. In this case the nucleon is exposed only to the nonzero modes that make up the sea, and correlators composed of quark propagators restricted to these modes reproduce ordinary hadron correlators at small quark masses.

We can assume that the strong interactions are largely confined to a box of about $(2.5 \,\textrm{fm})^4$, the `femto universe'. We take that as our reference volume, $V_0$. The exact size of $V_0$ does not matter for our final conclusion. The all-important question then is, what is the probability of finding a zero mode in the reference volume $V_0$? According to Fig.~\ref{fig2} the maximum of $n$ is at about $|Q| \approx \sqrt{\langle Q^2\rangle}$. Thus, on average, we can expect at most $R=\sqrt{\langle Q^2 \rangle}\,(V_0/V)$ zero modes in $V_0$. On a volume of $(10 \,\textrm{fm})^4$, which is about the largest volume the dipole moment has been computed on~\cite{Alexandrou:2020mds}, we find $R = 0.06$. On a volume of $(20 \,\textrm{fm})^4$ the ration shrinks by a factor of four to $R = 0.015$ corresponding to a probability of $1.5 \%$. We can exclude long-range correlations~\cite{Leutwyler:1992yt}. The final result is that the dipole moment vanishes proportional to
\begin{equation}
  |d_n| \, \propto \, \sqrt{\frac{\chi_t}{V}} \, |\theta| 
  \label{result}
\end{equation}
as the volume is taken to infinity. Corresponding estimates apply to any other hadronic observable, for example the CP violating pion-nucleon coupling constant $\bar{g}_{\pi NN}$~\cite{Crewther:1979pi}.

Alternatively, we could have expressed the topological charge by an integral of the Chern-Simons density over the boundary, making use of the fact that the Pontryagin density is a total derivative, which leads to the same conclusion.

\section{Theta vacuum: The general case}

We have assumed that the vacuum has not changed noticeably by the $\theta$ term, leaving the nucleon stable. This is justified for small values of $\theta$. To see what happens with larger $\theta$, I have to go a little further. The QCD vacuum is characterized by gluon and quark condensates. The benchmark for our considerations is the gluon condensate
\begin{equation}
  G = \frac{\alpha_s}{\pi} \langle G_{\mu\nu}^a G_{\mu\nu}^a\rangle \,.
\end{equation}
It is known to be independent of the scale parameter $\mu$~\cite{Kluberg-Stern:1974iel,Vainshtein:1981wh}, but depends on the scheme. In the gradient flow scheme~\cite{Schierholz:2024lge}
\begin{equation}
  G =  \frac{192}{\pi^2}\, \alpha(\mu)^2 \mu^4 \,,
\end{equation}
where $\alpha(\mu) = (4\pi/3)\, t^2 \langle E\rangle$ and $\mu^2 = 1/8t$. To keep $G$ constant, the running coupling must increase linearly with $1/\mu^2$ for small values of $\mu$,\vspace*{-0.25cm}
\begin{equation} 
  \alpha(\mu) = \frac{\Lambda^2}{\mu^2} \,.
  \label{alpha}
\end{equation}
While our earlier statements on the running coupling~\cite{Nakamura:2021meh} were based on numerical simulations, we have a proof of the relation (\ref{alpha}) now~\cite{Schierholz:2024lge}. This is sometimes referred as infrared slavery, the flip side of asymptotic freedom. It has been shown that (\ref{alpha}) leads to a confining, asyptotically linear static potential that reproduces the spectrum of heavy quark bound states~\cite{Richardson:1978bt,Nakamura:2021meh}. The result that confinement can be traced back to a nonvanishing gluon condensate is not new~\cite{Dosch:1987sk,Simonov:1987rn}.

\begin{figure}[h!]
  \vspace*{0.25cm}
  \begin{center}
    \includegraphics[width=6.5cm]{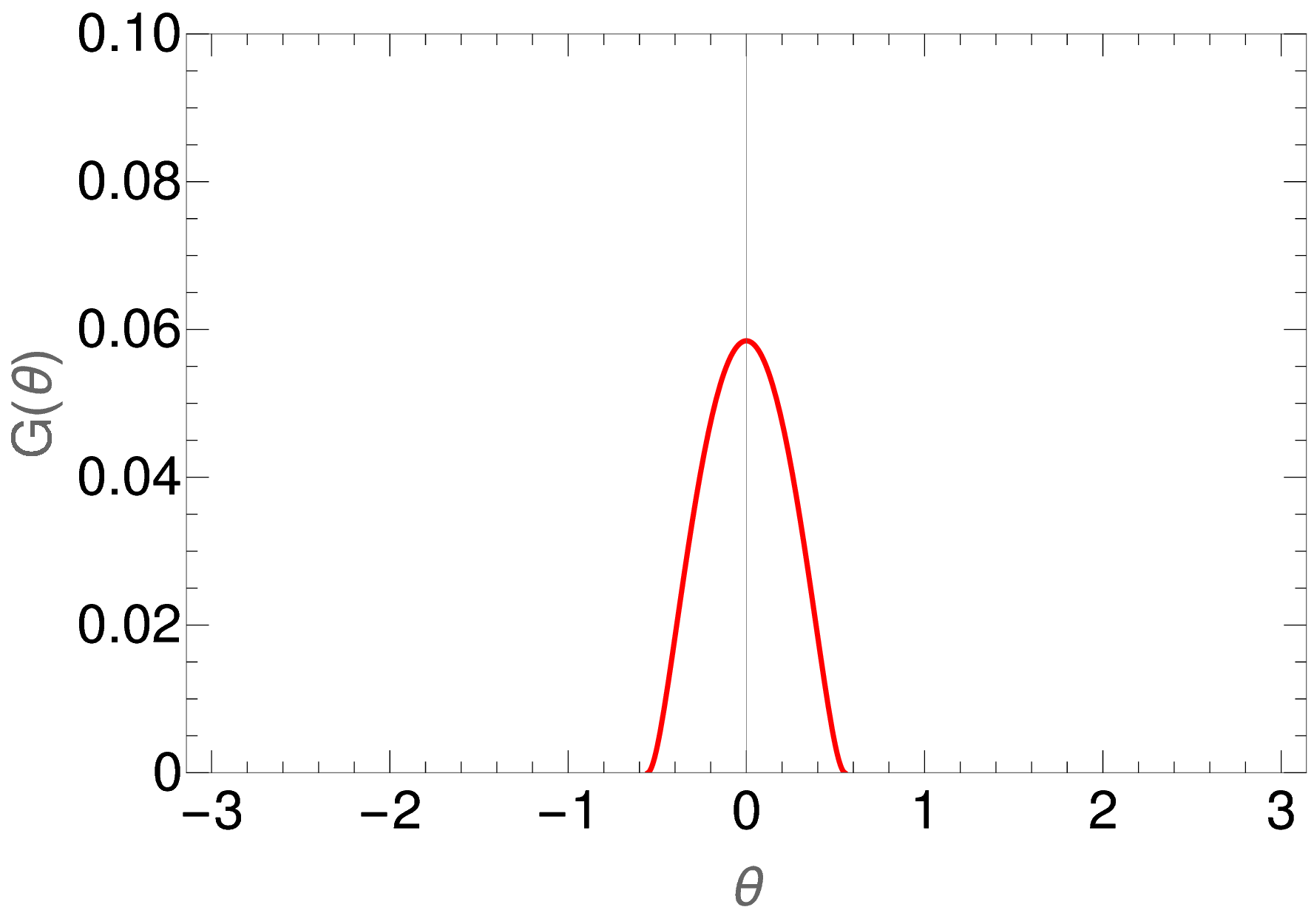}
  \end{center}
  \vspace*{-0.5cm}
  \caption{The gluon condensate in lattice units as a function of $\theta$ on the $24^4$ lattice at lattice spacing $a = 0.082\,\textrm{fm}$ for flow time $t/a^2= 60$.}
  \label{fig3}
\end{figure}

The gluon condensate in the $\theta$ vacuum, $G(\theta)$, is obtained from (\ref{op}). The determination follows the same pattern as the calculation of the running coupling constant~\cite{Nakamura:2021meh}. In Fig.~\ref{fig3} I show the result for the pure SU(3) gauge theory. We see that the gluon condensate vanishes outside a small region around $\theta = 0$. The curve is well approximated by a Gaussian. While $G(\theta)$ is independent of $\mu$ for $\theta = 0$ (by definition), its width decreases with decreasing $\mu$, similar to the running coupling. To keep $G(\theta)$ constant for $\theta >0$,  $\theta$ must be tuned, i.e.\ $\theta = \theta(\mu)$.

We conclude that the theory does not confine for $\theta > 0$. It has been suggested that QCD undergoes at least one phase transition within $|\theta| < \pi$~\cite{tHooft:1981bkw}, which we confirm. The gluon condensate is an even function of $\theta$. It arises from the real (CP-even) part of the partition function, which involves the density of topological charge squared, $Q^2/V$.

The transition to the deconfining region is governed by the screening radius $\lambda_S$, which was found to be $\lambda_S = 0.31/|\theta| \, \textrm{[fm]}$~\cite{Schierholz:2022wuc}. The nucleon is expected to disintegrate into quarks and gluons once the radius is smaller than the nucleon size. That might happen only for $|\theta| \gtrsim 0.4$. The situation is very similar to the ﬁnite temperature phase transition. While strictly conﬁnement is lost for $T > T_c$, a finite screening length, $\lambda_S \propto 1/(T-T_c)$, has been observed.

\section{The axion intrusion}

In the axion extension of QCD the $\theta$ term is augmented by the CP-conserving axion interaction~\cite{Peccei:1977hh},
\begin{equation}
 S_\theta \rightarrow S_\theta + S_\phi = \int d^4x \, \left[\frac{1}{2} \big(\partial_\mu \phi\big)^2 - i \left(\theta + \frac{\phi}{f}\right) \, P + L_\phi(\partial_\mu \phi\, J_\mu^5)\right]\,, \quad Q = \int d^4x \, P\,,
    \label{PQ}
\end{equation}
where $P$ is the Pontryagin density and $f$ is a scale factor. The Lagrangian $L_\phi$ describes the derivative couplings to the light quark axial vector currents. By construction the theory is invariant under global changes of the axion field, $\phi(x) \rightarrow \phi(x) + \delta$, called shift symmetry. Thus we may drop the $\theta$ term. The partition function (\ref{Z}) then becomes
\begin{equation}
  Z = \int\! \mathcal{D}[A_\mu] \int\! \mathcal{D}[\phi] \, \int\, [\cdots] \, \exp\Big\{\!-\!\int d^4x \,\Big[\frac{1}{2} \big(\partial_\mu \phi\big)^2 - \frac{i}{f}\, \phi\, P+ L_\phi\Big] - S_0\Big\}\,.
  \label{Z2}
\end{equation}
Physical quantities, like Green's functions, will not depend on the mean value of $\phi$. Writing
\begin{equation}
    \phi(x) = \phibar + \phitilde(x)\,, \quad \phibar \equiv \langle \phi\rangle =\frac{1}{V} \int d^4x\, \phi(x) \,,
    \label{cov}
\end{equation}
we may divide (\ref{Z2}) into an integral over $\phibar$ and a path integral over the field $\phitilde(x)$ orthogonal to $\phibar$. The integral over $\phibar$ can be carried out~\cite{Schierholz:2023hkx}, which results in a Kronecker delta function, $2\pi f \delta_{Q 0}$. Up to a factor we end up with the partion function
\begin{equation}
  Z = \int_{Q=0}\! \mathcal{D}[A_\mu] \int_{\langle \phi\rangle=0}\! \mathcal{D}[\phi] \, \int\, [\cdots] \, \exp\Big\{\!-\!\int d^4x \,\Big[\frac{1}{2} \big(\partial_\mu \phi\big)^2 + L_\phi\Big] - S_0\Big\}\,,
  \label{Z3}
\end{equation}
where we have changed $\phitilde$ to $\phi$ and moved the term $\pi P$, which can be reexpressed by $\partial_\mu \phi\, \Omega_\mu^{\scriptscriptstyle (0)}$,  into $L_\pi$~\cite{Schierholz:2023hkx}. $\Omega_\mu^{\scriptscriptstyle (0)}$ is the Chern-Simons density or 0-cochain. Restricting the integral to $\langle \phi\rangle =0$ is common practice in the calculation of the constraint effective potential. Continuum-like gauge fields split naturally into quantum mechanically disconnected sectors of fixed topological charge. Hence, the limitation to trivial topology is just a question of choice. The path integral over $\phi$ in (\ref{Z3}) can be sampled by a Gaussian white noise, which is an ergodic random process.

The limitation to the sector $Q = 0$ has far reaching consequences. First and foremost
\begin{equation}
  \chi_t = \frac{\langle Q^2\rangle}{V} = 0\,.
  \label{chi}
\end{equation}
The topological susceptibility enters into several observables. A milestone in establishing QCD is the Witten-Veneziano relation~\cite{Witten:1979vv,Veneziano:1979ec} \vspace*{-0.25cm}
\begin{equation}
  m_{\eta^\prime}^2 + m_{\eta}^2 - 2 m_K^2 = \frac{12}{f_\pi^2}\, \chi_t \,,
  \label{VW}
\end{equation}
where $\chi_t$ is to be taken in the limit of heavy quarks. Like no other formula it illustrates the role of topological charge in low-energy hadron physics. A quantity of great interest in axion phenomenology is the axion mass. It is given by~\cite{Peccei:2006as}
\begin{equation}
  m_a=\frac{\sqrt{\chi_t}}{f} \,.
  \label{axm}
\end{equation}
Axions are considered a candidate for dark matter. To estimate its mass in the post-inﬂation phase of the Universe, the topological susceptibility needs to be extrapolated to high temperatures. This was done in~\cite{Bonati:2015vqz,Borsanyi:2016ksw}. Assuming that dark matter is made up of axions, a mass 
\begin{equation}
  m_a=50(4)\, \mu\textrm{eV}
  \label{ma}
\end{equation}
has been suggested. Both results, (\ref{VW}) and (\ref{ma}), are invalidated by the result (\ref{chi}).

The absence of topological charge has severe consequences for the fermion sector. Most importantly, the theory has no anomaly. The action has a chiral $\textrm{U}_A(1)$ symmetry for small $q = u$ and $d$ quark masses, which thus is left unbroken. However, a nonvanishing chiral condensate $\langle \bar{q}q \rangle$ does not only break the $\textrm{SU(2)}_L \times \textrm{SU(2)}_R$ symmetry spontaneously but also the axial $\textrm{U}_A(1)$ symmetry. A spontaneously broken $\textrm{U}_A(1)$ symmetry would require an isoscalar Goldstone boson, the $\eta^\prime$, with a mass~\cite{Weinberg:1996kr} $m_{\eta^\prime} \leq \sqrt{3} \, m_\pi$. Its physical mass, however, is much heavier than the mass of the $\eta$ meson. This raises the question whether chiral symmetry is broken in the topologically trivial sector. Callan, Dashen and Gross~\cite{Callan:1977gz} suggested that chiral symmetry breaking is caused by the anomaly, which would predict \vspace*{-0.25cm}
\begin{equation} 
  2 m_q \langle \bar{q} q \rangle = \frac{\langle Q^2 \rangle}{V}
  \label{cc}
\end{equation}
for two light flavors. This has been confirmed by lattice simulations~\cite{Bruno:2014ova}. The answer would be no. Further support comes from studies of the chiral susceptibilities across the finite temperature phase transition~\cite{Aoki:2021qws}. The role of the $\textrm{U}_A(1)$ anomaly in QCD phenomenology, and what the loss of it would imply, has been summed up in~\cite{Shore:2007yn}, including the Gell-Mann--Oakes--Renner relation, meson decay constants, the Goldberger-Treiman relation, polarized structure functions and the proton spin.

\begin{figure}[h!]
  \vspace*{-0.5cm}
  \begin{center}
    \includegraphics[width=9cm]{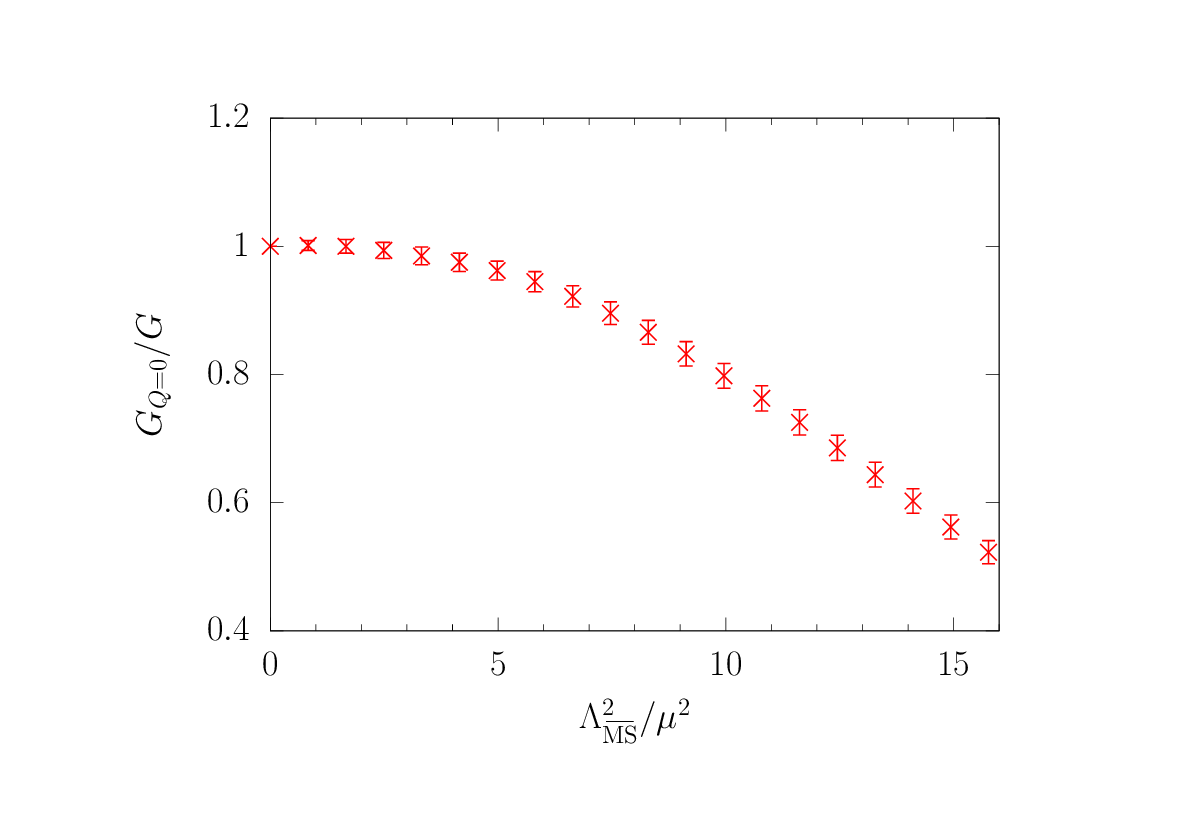}
  \end{center}
  \vspace*{-1.25cm}
  \caption{The ratio of gluon condensates $G_{Q=0}/G$ as a function of the scale parameter $\mu$ on the $32^4$ lattice at lattice spacing $a = 0.082\,\textrm{fm}$.}
  \label{fig4}
\end{figure}

There are indications that the Peccei-Quinn theory does not even confine. In Fig.~\ref{fig4} I show the gluon condensate in the topologically trivial sector, $G_{Q = 0}$ divided by $G$ as a function of the scale parameter $\mu$. We see that $G_{Q = 0}$ is no longer scale invariant, but tends to zero in the infrared. This does not come as a surprise, as the vacuum turns into a semi-classical ensemble of sign-coherent (anti-)instantons at large flow times~\cite{Nakamura:2021meh}. As a result, the running coupling is drastically reduced in the infrared, far from linear increase. This fits in with a vanishing quark condensate. \vspace*{-0.25cm}

\section{Conclusions} \vspace*{-0.25cm}

The QCD vacuum is specified by the $\theta$ angle. There are many speculations about the nature at nonvanishing $\theta$. CP violation is only one aspect of it. Ultimately, it is about the topological properties of the vacuum, which requires nonperturbative tools to address it, the lattice. It has been shown that the theory undergoes a transition to a deconfining phase at $|\theta| > 0$, with a screening length inversely proportional to $|\theta|$, and that CP is conserved in the transition region. The QCD vacuum is unique, and any extension of the Standard Model must not change its nonperturbative features. The axion extension is not such a model, within the set nonperturbative framework.

\end{document}